\documentclass[a4paper]{revtex4}
\usepackage[top=1in, bottom=1in, left=1in, right=1in]{geometry}
\usepackage{amsmath}
\usepackage{amssymb}
\usepackage{amsfonts}
\usepackage{graphicx,bm}
\usepackage{dcolumn}
\usepackage{epsfig}
\usepackage[colorlinks,linkcolor=blue,citecolor=blue,urlcolor=blue]{hyperref}

\def \lleq {\lower0.9ex\hbox{ $\buildrel < \over \sim$} ~}
\def \ggeq {\lower0.9ex\hbox{ $\buildrel > \over \sim$} ~}

\def \beq  {\begin{equation}}
\def \eeq  {\end{equation}}
\def \ber  {\begin{eqnarray}}
\def \eer  {\end{eqnarray}}

\newcommand{\be}{\begin{equation}}
\newcommand{\ee}{\end{equation}}
\newcommand{\ba}{\begin{eqnarray}}
\newcommand{\ea}{\end{eqnarray}}
\newcommand{\bea}{\begin{eqnarray*}}
\newcommand{\eea}{\end{eqnarray*}}

\begin{document}

\title{Observational constraints on the generalized $\alpha$ attractor model }
\author{M. Shahalam$^1$\thanks{%
E-mail address: shahalam@zjut.edu.cn}, Ratbay Myrzakulov$^2$\thanks{%
E-mail address: rmyrzakulov@gmail.com}, Shynaray Myrzakul$^3$\thanks{%
E-mail address: shynaray1981@gmail.com},  Anzhong Wang$^{1,4}$\thanks{%
E-mail address: Anzhong_Wang@baylor.edu}}
\affiliation{$^{1}$Institute for Advanced Physics $\&$ Mathematics, Zhejiang University of Technology, Hangzhou, China\\
$^2$Eurasian International Center for Theoretical Physics, Department of General and Theoretical Physics, Eurasian National
University, Astana, Kazakhstan\\
$^3$Department of Theoretical and Nuclear Physics, Al-Farabi Kazakh National University, Al-Farabi Almaty, Kazakhstan
\\
$^4$GCAP-CASPER, Department of Physics, Baylor University, Waco, Texas, USA}
\begin{abstract}
We study the generalized $\alpha$ attractor model in the context of the late time cosmic acceleration. The model  interpolates between the scaling freezing and thawing dark 
energy models. In the slow roll region, the original potential is modified whereas the modification ceases in the asymptotic region and the effective potential behaves 
as the quadratic one. In our setting, the field rolls slowly around the present epoch and mimics the scaling behavior in the future. We obtain observational constraints on the model parameters by using an integrated data base (SN+Hubble+BAO+CMB).
\end{abstract}

\maketitle

\section{Introduction}
\label{intro}
The past two decades of tremendous activities in observational cosmology
motivated theorists to consider various theories of  
gravity. This leads to a plethora of cosmological models at the theoretical
ground. The observations of Type Ia supernovae by the Supernova Cosmology
Project and the High-Z Supernova Search Team provided strong evidence  about
the currently accelerated expansion of the Universe (against the presumed decelerating
expansion caused by gravity). Baryon acoustic oscillations and
 results from the clustering of galaxies yielded  more
confirmatory evidences to this acceleration. Nowadays the concept of the late time cosmic acceleration  has become a fundamental ingredient for theorists in modeling  the Universe. 
The idea of the late time cosmic acceleration is mainly attributed to the
presence of some mysterious entity commonly referred to as Dark Energy \cite{sami,sahni,sami1,sami2,sami3,sami4,sami5,sami6}. Although there have been developed several 
ideas to explain the cosmic acceleration, the theory of Dark Energy is much more successful to explaining
various phenomena. On the other hand,  primordial inflation has taken a
special status in explaining the origin of the anisotropies in the cosmic
microwave background radiation (CMBR) and the formation of the large scale
structure. The origin of both the early and late time inflation still
represents a great theoretical puzzle which motivates theorists to invoke
scalar fields to explain the two inflationary phases simultaneously. During
the past three and half decades, a wide variety of inflationary models
have been proposed,  among which \textquotedblleft cosmological
attractors\textquotedblright\ was discovered very recently. These belong to a very
broad class which incorporates the conformal attractors \cite{conformal,conformal1},
alpha attractors \cite{alpha,alpha1,alpha2,alpha3,alpha4},  and also includes scalar field
cosmological models such as the Starobinsky model \cite{staro, staro1,staro2,staro3,staro4}, the chaotic
inflation in super-gravity (GL model) \cite{GL,GL1,GL2}, Higgs inflation \cite{higgs,higgs1,higgs2,higgs3,higgs4,higgs5,higgs6,higgs7,higgs8,higgs9} and axion monodromy inflation
 \cite{mono,mono1,mono2,mono3,mono4,mono5,mono6,mono7}. All these have a mysterious fact in the context of recently released results obtained by WMAP and Planck data 
 \cite{planck,planck1,planck2,planck3} that they provide the very similar cosmological
predictions, although they have different origins. For conformal attractors, they predict that, for a large number of $e$-folds $N$, the spectral index and tensor-to-scalar
ratio are given by $n_{s}=1-2/N$ ; $r=12/N^{2}$. For $N\sim 60$, these
predictions are $n_{s}\sim 0.967,r\sim 0.003$, while for $N\sim 50$, $%
n_{s}\sim 0.96,r\sim 0.005$,  which are in very good agreement with WMAP and
Planck data. For $\alpha$ attractors, the slow roll parameters $n_s$ and $r$ can be written as   $n_s=1-2/N,  r=12 \alpha/N^2$ for small $\alpha$, and  $n_s= 1-2/N, r=12 \alpha/(N(N+3 \alpha /2))$ for large $\alpha$,
where $N$ is the number of e-folds between the end of inflation and  horizon-crossing, and its numerical value lies in the range $50 \leq N \leq 60$. These models can be used not only for inflation but also   for
 the late time cosmic acceleration \cite{super,super1,super2,super3,super4,super5,super6,super7}. In this paper, we shall study the generalized $\alpha$ attractor model in the
 context of the late time cosmic acceleration.

The paper is organized as follows. In Section \ref{model}, we present the basics of the $\alpha$ attractor models. Section \ref{evol} displays the evolution equations of the scalar field in the autonomous form and 
addresses the issue of the cosmological attractor.  In Section \ref{DA}, we use the joint data to carry out observational analyses and obtain constraints on the model parameters. Our results are summarized in Section \ref{conclusion}.
\vskip 1cm
\section{$\alpha$ - Models}
\label{model}

Here, we focus on the minimally coupled alpha attractors which are of more
interest as they play an important role in cosmology to investigate the
dark energy properties and  to evaluate if these models can explain the
late time cosmic acceleration. In the Einstein frame, the Lagrangian density of the
alpha attractor model takes the form,
\begin{equation}
\mathcal{L}=\sqrt{-g}\left[ \frac{1}{2} M_p^2R-\frac{\alpha }{\left( 1-\frac{\varphi ^{2}}{6}%
\right) ^{2}}\frac{\left( \partial \varphi \right) ^{2}}{2}-\alpha
f^{2}\left( \frac{\varphi }{\sqrt{6}}\right) \right]
\label{eq:lag}
\end{equation}%
with  a real scalar field $\varphi$. Despite of the changes in the inflaton
potential and an arbitrary function $f\left( \varphi \right)$,  this class of
models has similar striking results for the primordial scalar perturbation tilt
and tensor-to-scalar ratio. For $\alpha =1$, the Starobinsky model
is recovered. We would like to note that the kinetic and potential energies
have the same overall coefficient $\alpha$. The kinetic term is
not canonical, but  can be so through a field redefinition $\phi =\sqrt{%
6\alpha }\tanh ^{-1}\left( \frac{\varphi }{\sqrt{6}}\right) $ with  $V\left( \phi \right) =\alpha f^{2}\left( \tanh \left( 
\frac{\phi }{\sqrt{6\alpha }}\right) \right) $. We are interested in the dynamics of the scalar field and the evolution of its equation of state, and see whether it explains the late-time cosmic acceleration.

Two functional  forms of $f$ have been employed for inflation, 
\begin{eqnarray}
f &=& c~  \tanh \left( \frac{\phi }{\sqrt{6\alpha }}\right),\\
\text{and} \qquad f &=& c~  \frac{\tanh \left( \frac{\phi }{\sqrt{6\alpha }}\right)}{1+\tanh \left( \frac{\phi }{\sqrt{6\alpha }}\right)}
\end{eqnarray}
The first one is known as the T model and  the second one is  the Starobinsky model ($\alpha$=1). Both models are identical at small $\phi$ and behave as the quadratic potential near the origin.

\begin{figure}[tbp]
\begin{center}
\begin{tabular}{ccc}
{\includegraphics[width=2.2in,height=2.2in,angle=0]{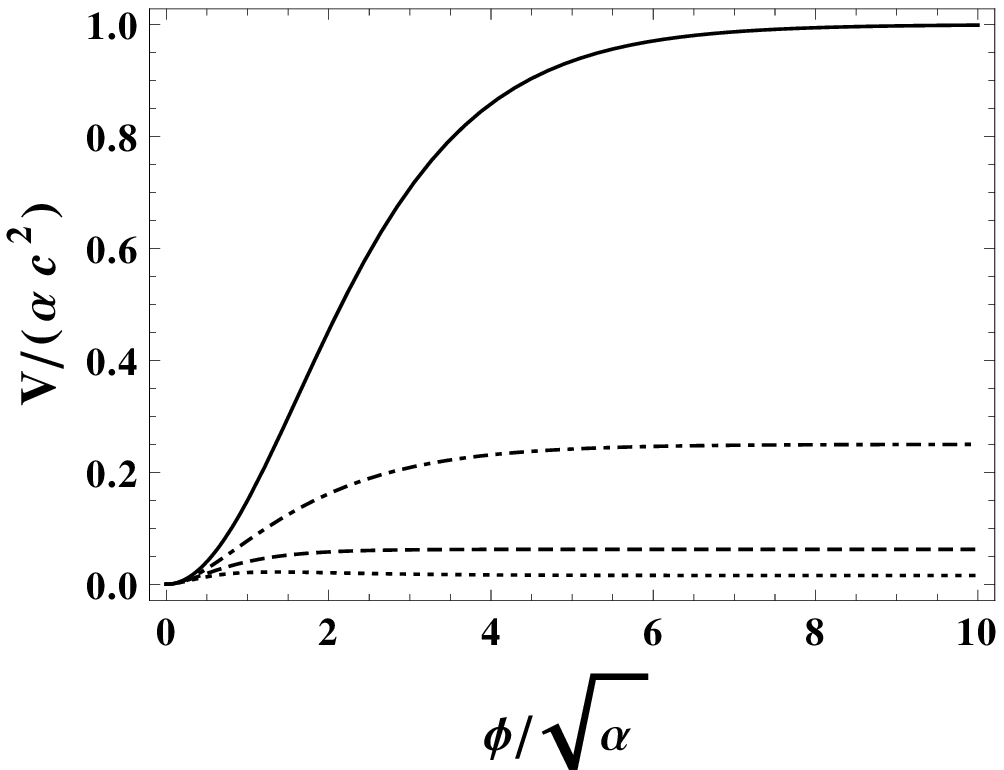}} &
{\includegraphics[width=2.2in,height=2.2in,angle=0]{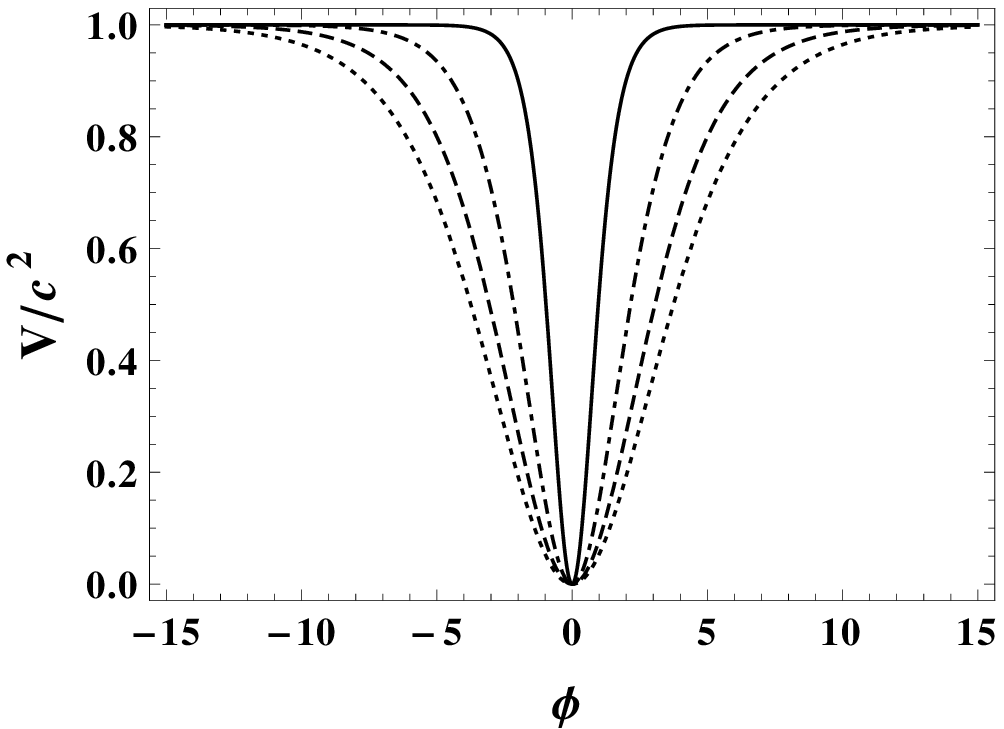}} 
\end{tabular}
\end{center}
\caption{In the left panel, the potential (\ref{pot}) of the generalized $\alpha$ model is plotted against $\phi/\sqrt{\alpha}$, for $n=0, 1, 2, 3$ (from top to bottom). For large field values ($\phi \gg \sqrt{\alpha}$ ), the potential becomes constant and for small field values ($\phi \ll \sqrt{\alpha}$ ), it behaves as a quadratic potential. The right panel shows the evolution of the potential versus the field $\phi$ for different values of $\alpha$. The dot, dashed, dot-dashed and solid lines represent the values of 
$\alpha=3, 2, 1, 0.2$, respectively. The smaller $\alpha$ corresponds to a more narrow minimum of the potentials. This panel is plotted only for $n=0$. In both  panels, the potential and field are shown in units of $ c^2$ and Planck, respectively.}
\label{figpot}
\end{figure}

We consider  a generalized $\alpha$ model given by \cite{linder15}
\begin{eqnarray}
f &=& c~  \frac{\tanh \left( \frac{\phi }{\sqrt{6\alpha }}\right)}{\left(1+\tanh \left ( \frac{\phi }{\sqrt{6\alpha }}\right)\right)^n}
\end{eqnarray}
where $c$ is a constant, $\alpha$ is a parameter, and $n$ is a number which can take the values $n = 0,1,2,3, ...$ Therefore, the potential of the generalized  $\alpha$ model becomes  
\begin{eqnarray}
V(\phi) &=& \alpha~ c^2~  \frac{\tanh \left( \frac{\phi }{\sqrt{6\alpha }}\right)^2}{\left( 1+\tanh \left( \frac{\phi }{\sqrt{6\alpha }}\right) \right)^{2n}}
\label{pot}
\end{eqnarray}
As $\phi \rightarrow \infty$, the potential becomes constant (flatten), and as $\phi \rightarrow 0$, it behaves as a quadratic potential (see Figure \ref{figpot}).
%%%%%%%%%
\begin{figure}[tbp]
\begin{center}
\begin{tabular}{ccc}
{\includegraphics[width=1.6in,height=1.6in,angle=0]{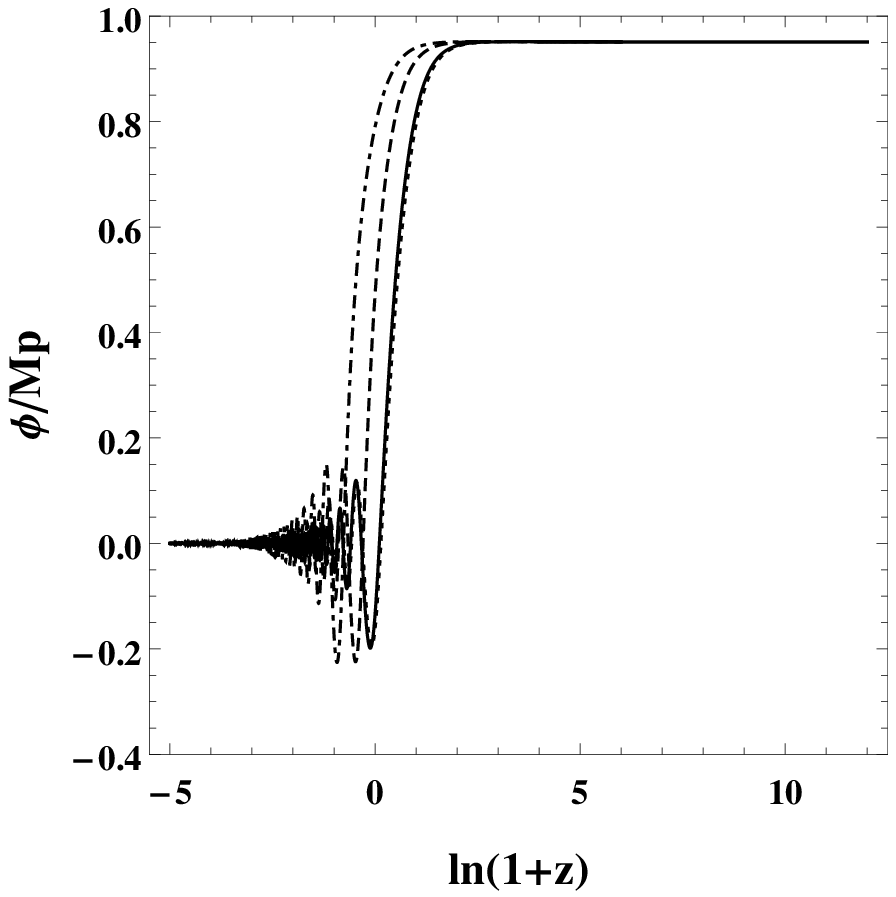}} &
{\includegraphics[width=1.6in,height=1.6in,angle=0]{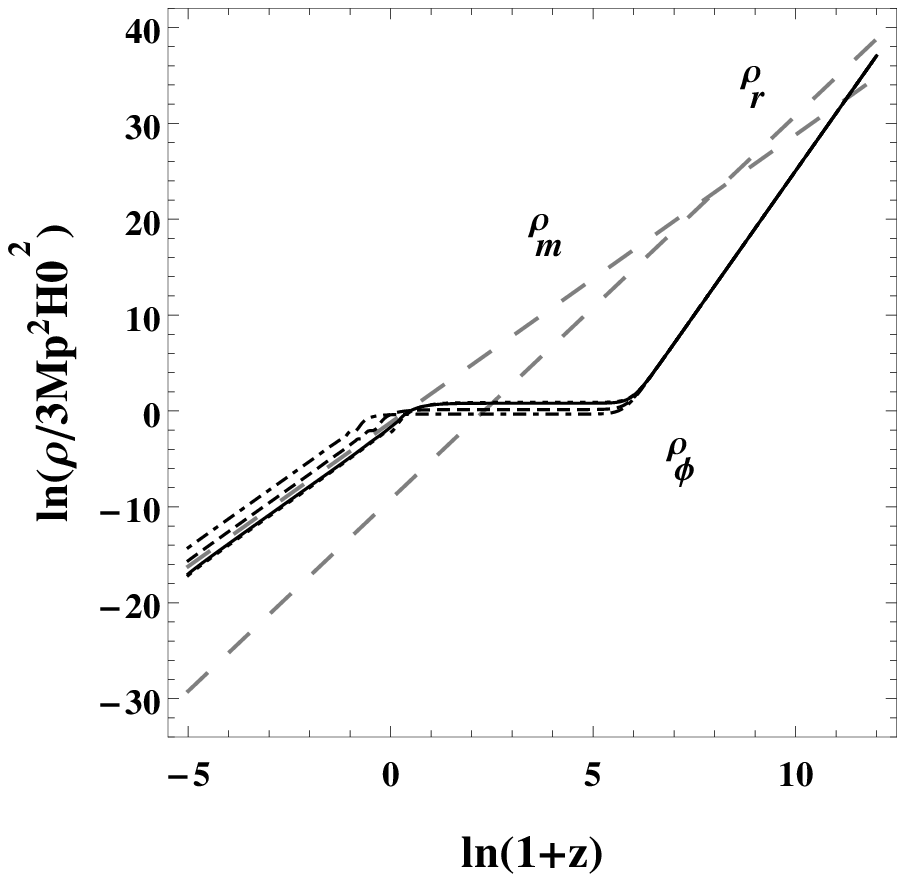}} &
{\includegraphics[width=1.6in,height=1.6in,angle=0]{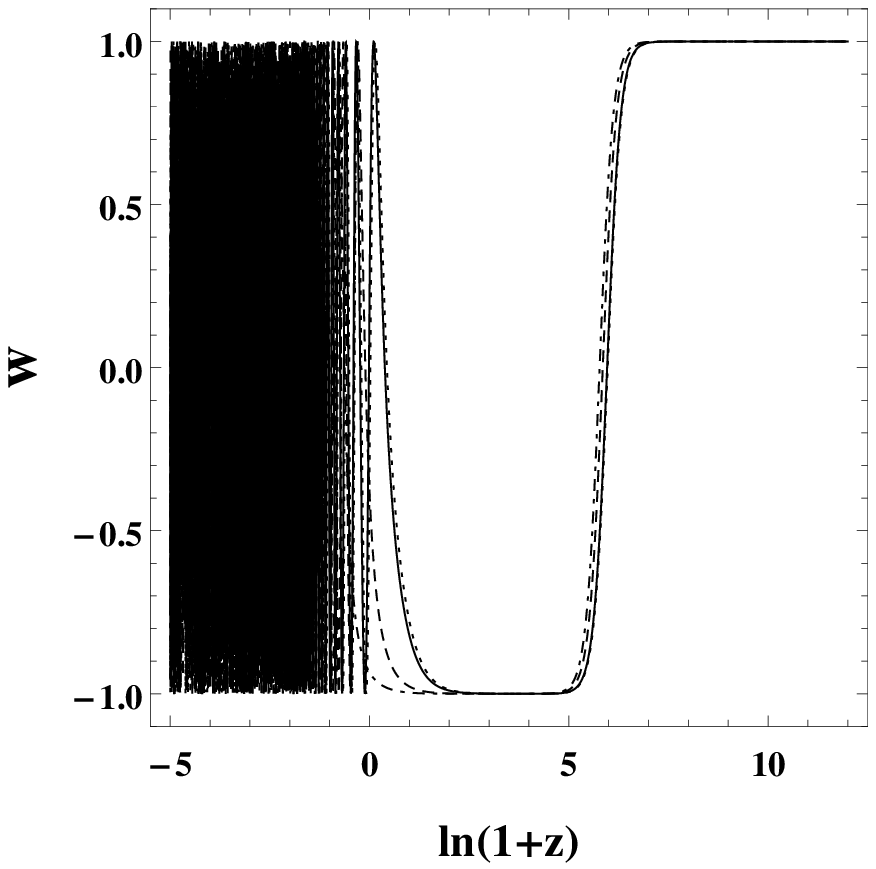}} 
 \\
{\includegraphics[width=1.6in,height=1.6in,angle=0]{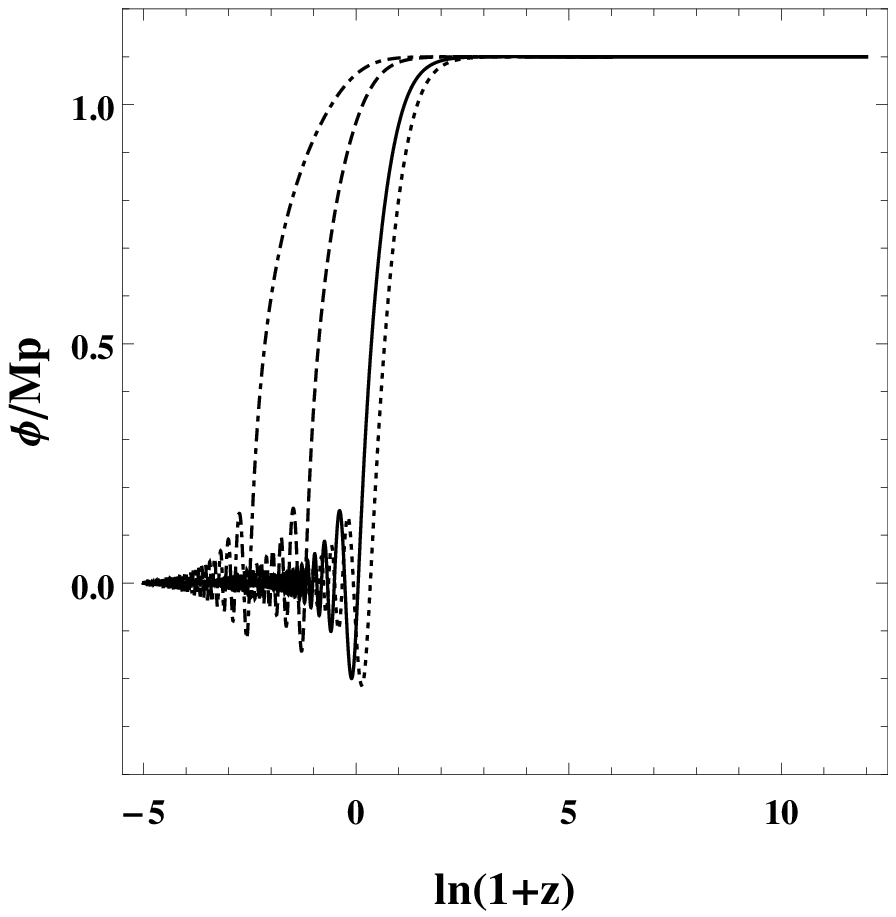}} &
{\includegraphics[width=1.6in,height=1.6in,angle=0]{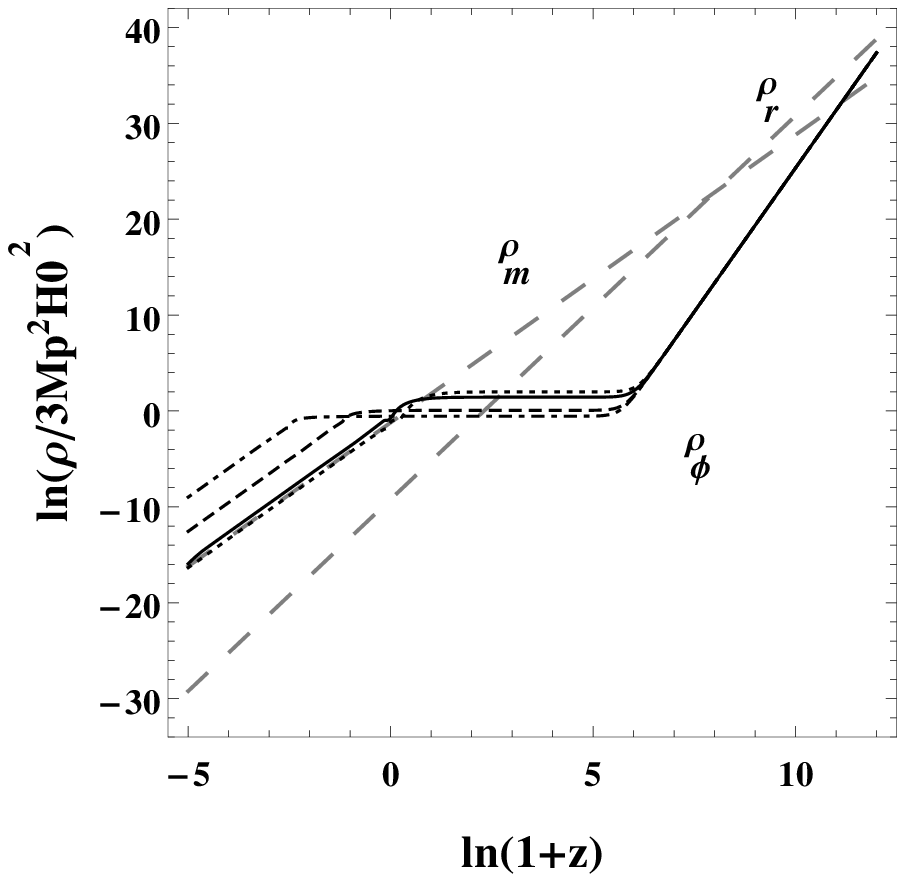}} & 
{\includegraphics[width=1.6in,height=1.6in,angle=0]{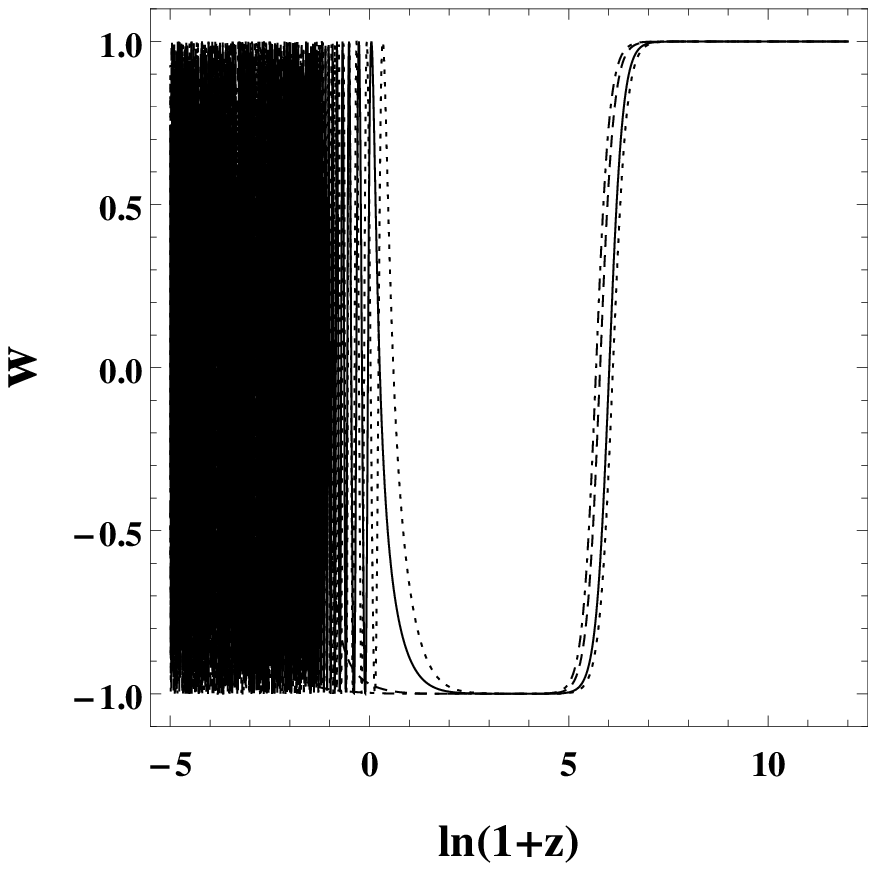}} 
\end{tabular}
\end{center}
\caption{The figure exhibits the evolution of the field $\phi$, energy density $\rho$ and equation of state $w$ versus the redshift $z$. The dot-dashed, dashed, solid,  and dotted lines correspond to the evolution of $\phi$, $\rho_{\phi}$ and $w$ for   $\alpha=0.05, 0.1, 1,  10$, respectively. The big dashed ($-$) lines represent the evolution of the energy densities of matter and radiation. In the upper and lower left panels, the field evolves from plateau region, as the time passes, it approaches to origin and gives rise to oscillating behavior. The upper and lower middle panels show that, initially the field energy density $\rho_{\phi}$ is sub-dominant and remains so for almost all of the period of the evolution. At late times, $\rho_{\phi}$ catches up with the energy density of the background (big dashed lines) and eventually overtakes it. Around the present epoch, $\rho_{\phi}$ lies in the thawing region as the field freezes due to a large Hubble damping, and in future it scales with the background. The upper and lower right panels show the evolution of the equation of state $w$. At the present epoch, it acts as a thawing, and in future as the field approaches to origin, the equation of state oscillates between $+1$ and $-1$, and the system spends most of the time around $w=\pm 1$. The upper panels are plotted for $\Omega_{0m}=0.3$, $c=7 M_p H_0$ and $n=0$ whereas the lower panels are for $\Omega_{0m}=0.3$, $c=12 M_p H_0$ and $n=1$.}
\label{figrho}
\end{figure}
%%%%%%%
\begin{figure}[tbp]
\begin{center}
\begin{tabular}{c}
{\includegraphics[width=2.1in,height=2.1in,angle=0]{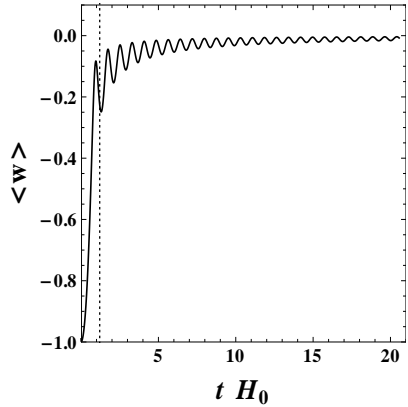}}
\end{tabular}
\end{center}
\caption{The figure displays the average equation of state versus  the cosmic time. In the future, the potential behaves as a quadratic potential whose average equation of state is $< w > =0$ at the attractor point, as shown in the figure, which is consistent with the analytical solution (\ref{eq:wavg}). The vertical dotted line represents the present epoch. }
\label{figwavg}
\end{figure}
%%%%%%%%%
\begin{figure}[tbp]
\begin{center}
\begin{tabular}{ccc}
%{\includegraphics[width=2.1in,height=2.1in,angle=0]{amn0.eps}} &
%{\includegraphics[width=2.1in,height=2.1in,angle=0]{acn0.eps}} &
%{\includegraphics[width=2.1in,height=2.1in,angle=0]{mcn0.eps}} 
%\\
{\includegraphics[width=1.6in,height=1.6in,angle=0]{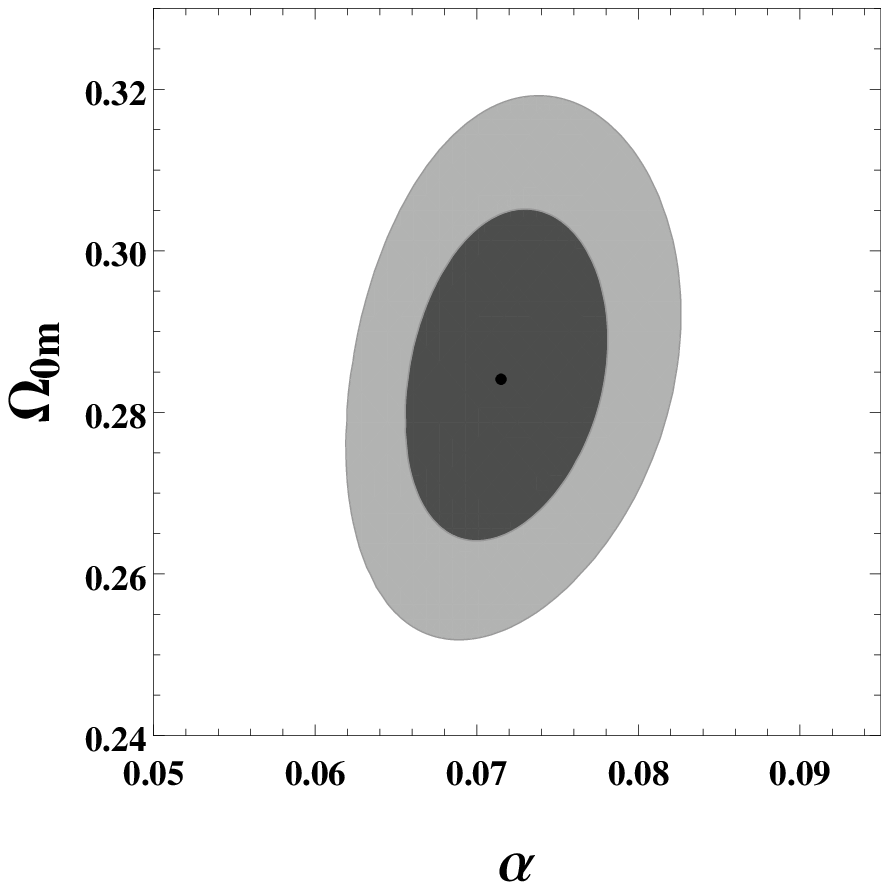}} &
{\includegraphics[width=1.6in,height=1.6in,angle=0]{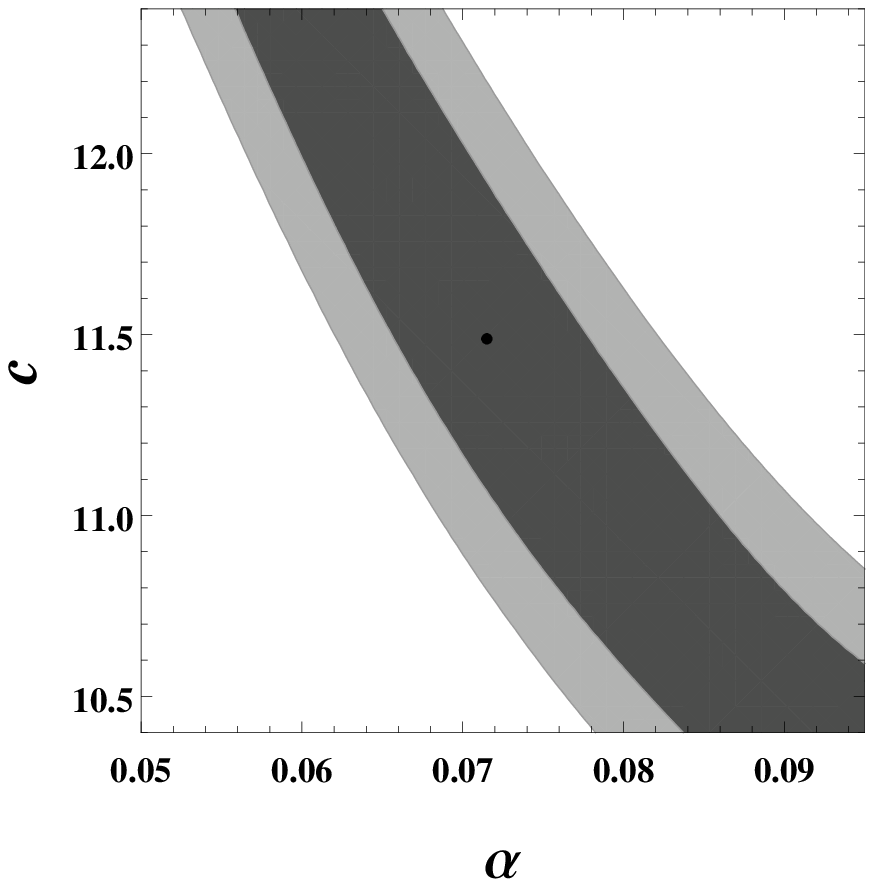}} & 
{\includegraphics[width=1.6in,height=1.6in,angle=0]{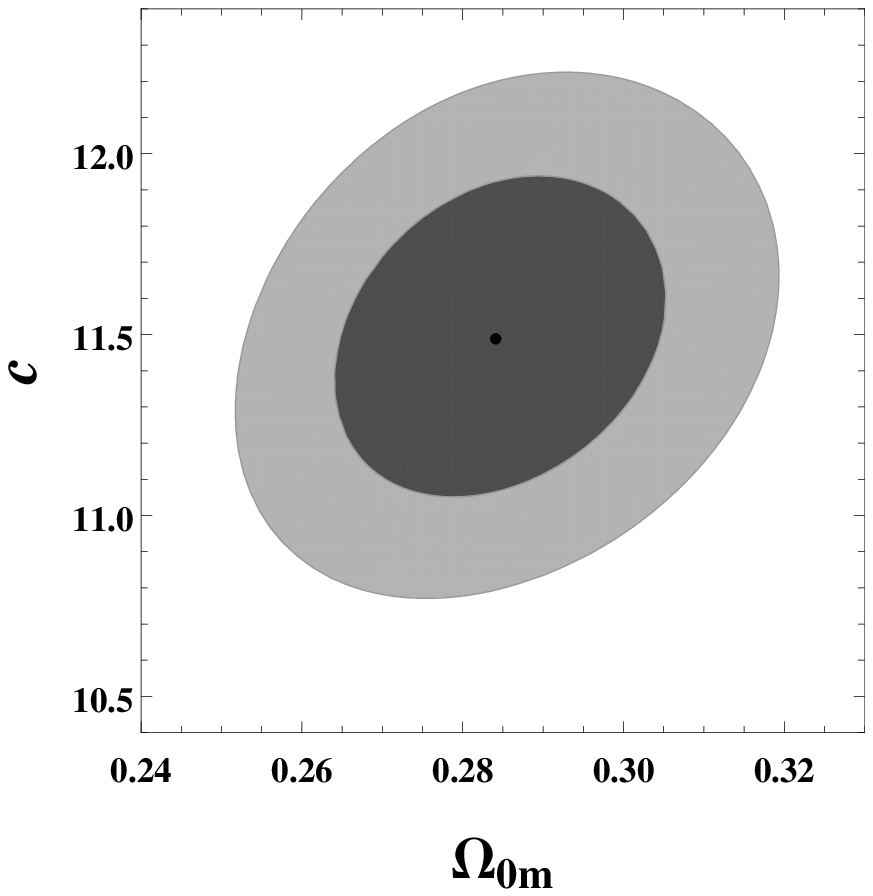}} 
\\
{\includegraphics[width=1.6in,height=1.6in,angle=0]{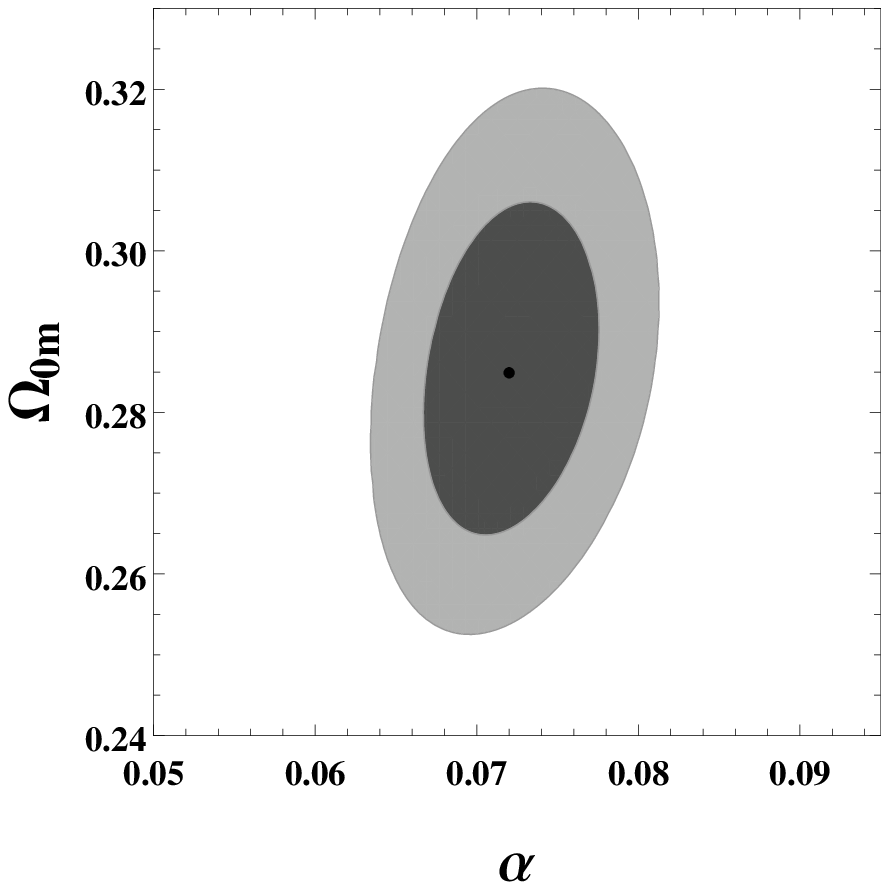}} &
{\includegraphics[width=1.6in,height=1.6in,angle=0]{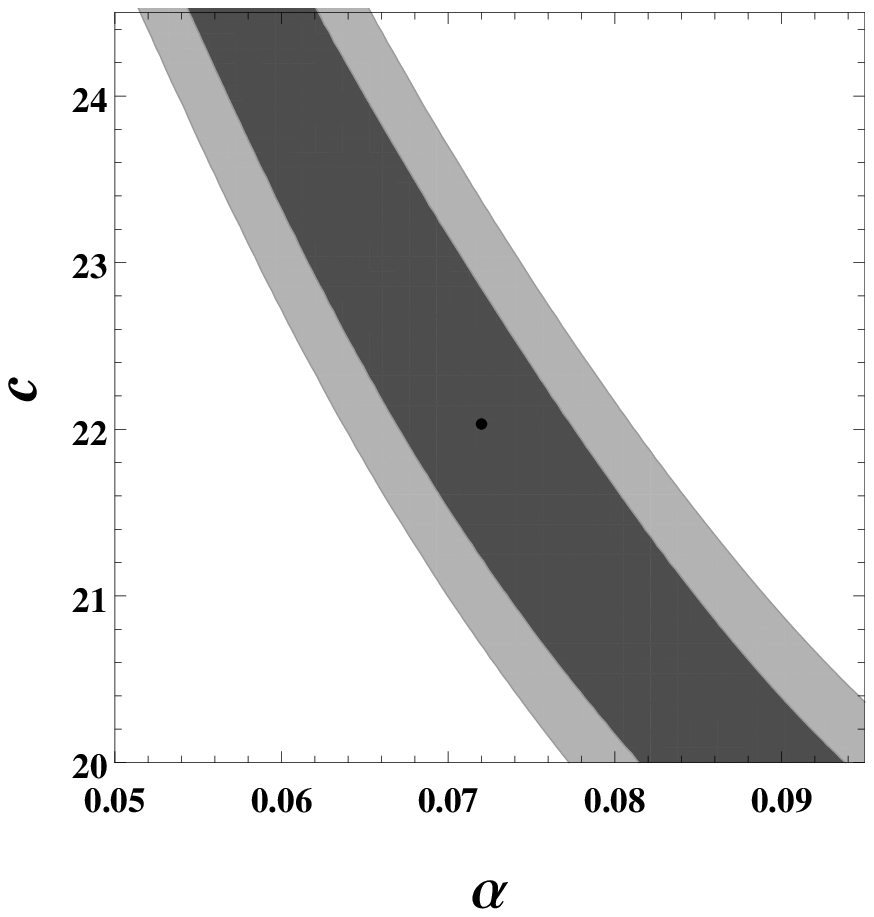}} & 
{\includegraphics[width=1.6in,height=1.6in,angle=0]{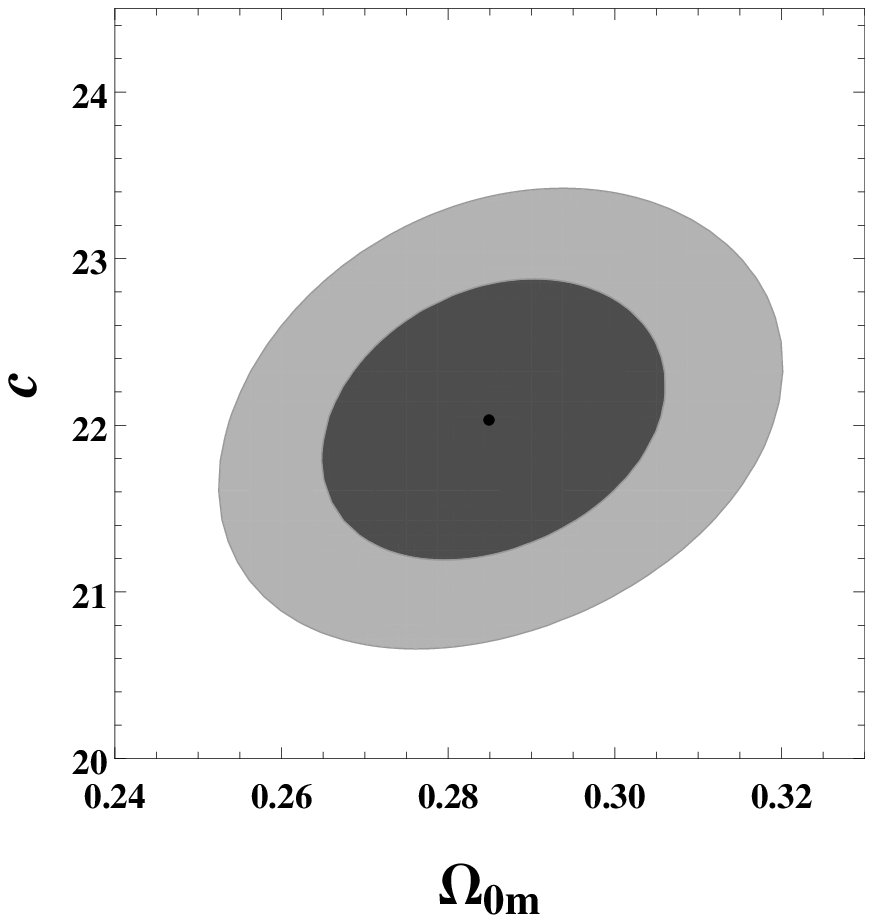}} 
\end{tabular}
\end{center}
\caption{Top and bottom panels show 1$\sigma$ (dark shaded) and 2$\sigma$ (light shaded) likelihood contours for $n=1, 2$, respectively. We have used the joint data (SN + Hubble + BAO + CMB) to carry out the data analysis. 
The black dots represent the best-fitting values of the parameters.}
\label{figobs}
\end{figure}
%%%%%%%
%%%%%%%
\section{Evolution equations and attractors}
\label{evol}

Caldwell and Linder \cite{cald} revealed that the scalar field models can be divided into two categories: the fast roll (freezing) and slow roll (thawing) models. A freezing model is such that, during the matter/radiation era, the field mimics the background and remains sub-dominant. Only at late times, the field exits to late time cosmic acceleration. The freezing models remain independent for a wide range of initial conditions. This class corresponds to tracking freezing model. The another sub-class of freezing models is associated with the scaling solutions \cite{wand}. In this case, the energy density of the scalar field scales with the background energy density during most of the matter era.

In contrast, the thawing models are alike to inflaton that derives the acceleration of the Universe at early epoch. In the thawing models, the scalar field is initially frozen due to a large Hubble damping and behaves as a cosmological constant with $w \approx -1$. At late times the Hubble damping decreases and the scalar field slowly thaws from the frozen state and deviates from behavior of  the cosmological constant. Thawing models are much sensitive to initial conditions. Some of the models of this class have been studied in  \cite{msa1,msa2}. The thawing, tracker and scaling models have been discussed in \cite{shinji}.

The $\alpha$ attractor models seem to combine these two classes in one way. We consider the potential given by equation (\ref{pot}), which has  the  following asymptotic form
\begin{eqnarray}
\qquad V(\phi) \approx \alpha~ c^2~  2^{-2n} \left(1-2(2-n) e^{-\frac{2 \phi}{\sqrt{6\alpha }}}\right),  ~~~~~~~~~~~~~~~~~~~    \phi \gg \sqrt{\alpha }
\end{eqnarray}
This is an uplifted exponential potential and studied as the inflationary models \cite{GL,GL1,GL2}. The exponential potential falls in the freezing class and approaches to a cosmological constant.
\begin{eqnarray}
\qquad\qquad V(\phi) \approx   \frac{c^2}{6 \alpha} \phi^2 ,  ~~~~~~~~~~~~~~~~~~~~~~~~~~~~~    \phi \ll \sqrt{\alpha }
\end{eqnarray}
This is the quadratic potential, falls in the thawing class  and deviates from the cosmological constant. The analytical solution of the average equation of state for the power-law potential $V(\phi) \propto \phi^{2p}$ is given by
\begin{eqnarray}
< w > &=& \frac{p-1}{p+1}
\label{eq:wavg}
\end{eqnarray}
For the quadratic potential $p=1$, therefore, in this case, as $\phi$ approaches to the origin, the time average equation of state during the oscillations is given by $< w > = 0$.
%To have a viable thermal history of the Universe, we need to have $\Omega_{\phi}= \frac{3(1+w_B)}{(2/\sqrt{6 \alpha})^2}$ = constant $\leq 0.01$ \cite{planck} during the radiation-dominated era, which implies $\alpha \leq 0.0017$ (here $w_B$ represents the background equation of state).

The equations of motion are obtained by varying the Lagrangian density (\ref{eq:lag}) with equation (\ref{pot}) have the following forms 
\begin{eqnarray}
\frac{\dot{a}^2(t)}{a^2(t)} &=& \frac{1}{3 M_{pl}^2}   \left( \rho_r  + \rho_m +\rho_{\phi} \right)
\label{a}
\end{eqnarray}
\begin{eqnarray}
\ddot{\phi} +3 \frac{\dot{a}}{a} \dot{\phi}+ V'(\phi)=0
\label{phi}
\end{eqnarray}
where a prime ($'$) denotes the derivative with respect to $\phi$. In the following discussions  we shall use the dimensionless variables
\begin{equation}
Y_1={\phi \over M_p},\quad Y_2={\dot{\phi} \over M_{p} H_{0}},\quad {\cal V}={ V(Y_1) \over M_p^2 H_0^2}.
\end{equation}
Using these new dimensionless variables, we can cast equations (\ref{a}) and (\ref{phi}) as a system of the first-order equations 
\begin{equation}
Y_1'=\frac{ Y_2}{h(Y_1,Y_2)} \,
\label{evol1d}
\end{equation}
\begin{equation}
Y_2'= -3Y_2-{1 \over h(Y_1, Y_2)}\Big[{d {\cal V}(Y_1) \over dY_1} \Big]
\label{evol2d}
\end{equation}
The prime ($'$) denotes the derivative with respect to $\ln(a)$, and the function
$ h(Y_1, Y_2)$ is given by,
\begin{equation}
 h(Y_1, Y_2)=\sqrt{\left[{Y_2^2 \over 6}+ {{\cal V}(Y_1) \over 3} +{\Omega_{0m} e^{-3a}} +{\Omega_{0r} e^{-4a}}  \right]} \label{hubble}
\end{equation}
Here, $\Omega_{0r}$ and $\Omega_{0m}$ are the energy density parameters of radiation and matter, respectively at the present epoch. We solve the evolution equations (\ref{evol1d}) and (\ref{evol2d}) numerically, and the results are shown in Figures \ref{figpot} - \ref{figobs}. Figure \ref{figpot} exhibits the behavior of the potential (\ref{pot}) versus the scalar  field $\phi$, in units of $c^2$ and Planck, respectively. The left panel is plotted for $n=0,1,2, 3$. For large field values, the potential flattens to an uplifted plateau, while at small values it looks  like the quadratic potential. The right panel is plotted for $n=0$ with various values of $\alpha$. The smaller values of $\alpha$ correspond to a more narrow minimum of the potential.

As the field  evolving from the flatten region approaches the origin, the energy density $\rho_{\phi}$ undershoots the background and begins as thawing dark energy along with thawing behavior as the field freezes due to a large Hubble damping, and  it lies in the thawing region even at the present epoch. However,  in the future it switches over and converts to the scaling freezing behavior (see Figure \ref{figrho}).

The upper and lower right panels of Figure \ref{figrho} show that the field oscillates most time between $w= \pm 1$, and correspondingly the average equation of state is $ < w > =0$ at the attractor point (see Figure \ref{figwavg}), which is consistent with the analytical result (\ref{eq:wavg}) for a quadratic potential.  

Figure \ref{figobs} demonstrates the 1$\sigma$ (dark shaded) and 2$\sigma$ (light shaded) likelihood contours in the $\alpha - \Omega_{0m}$, $\alpha - c$ and $\Omega_{0m} - c$ planes with two different values of $n$, for the  generalized $\alpha$ attractor model. The joint data (SN+Hubble+BAO+CMB) have been used to carry out the observational analysis. The best-fitting values of the model parameters for two distinct values of $n$ are shown in Table \ref{bestt}.
%%%%%%%%%
\begin{table}
\caption{The Table represents the best-fitting values of the model parameters for two values of $n$.}
\begin{center}
\label{bestt}
\begin{tabular}{cccc}
\hline\hline
~~$n$ & ~~~~~~~~$\alpha$ &~~~~~~~~~~ $\Omega_{0m}$ & ~~~~~~ $c$ \\
\tableline
%\\
%~~0&~~~~~~~~    0.0500&~~~~~~~~~~  0.2800&~~~~~~~~ 7.2721\\
\\
~~1&~~~~~~~~    0.0715&~~~~~~~~~~  0.2841&~~~~~~~~  11.4884\\
\\
~~2&~~~~~~~~    0.0720&~~~~~~~~~~  0.2849&~~~~~~~~ 22.0320\\
\\
\hline\hline
\end{tabular}
\end{center}
\end{table}
%%%%%%%

\begin{figure}[tbp]
\begin{center}
\begin{tabular}{cc}
{\includegraphics[width=1.6in,height=1.6in,angle=0]{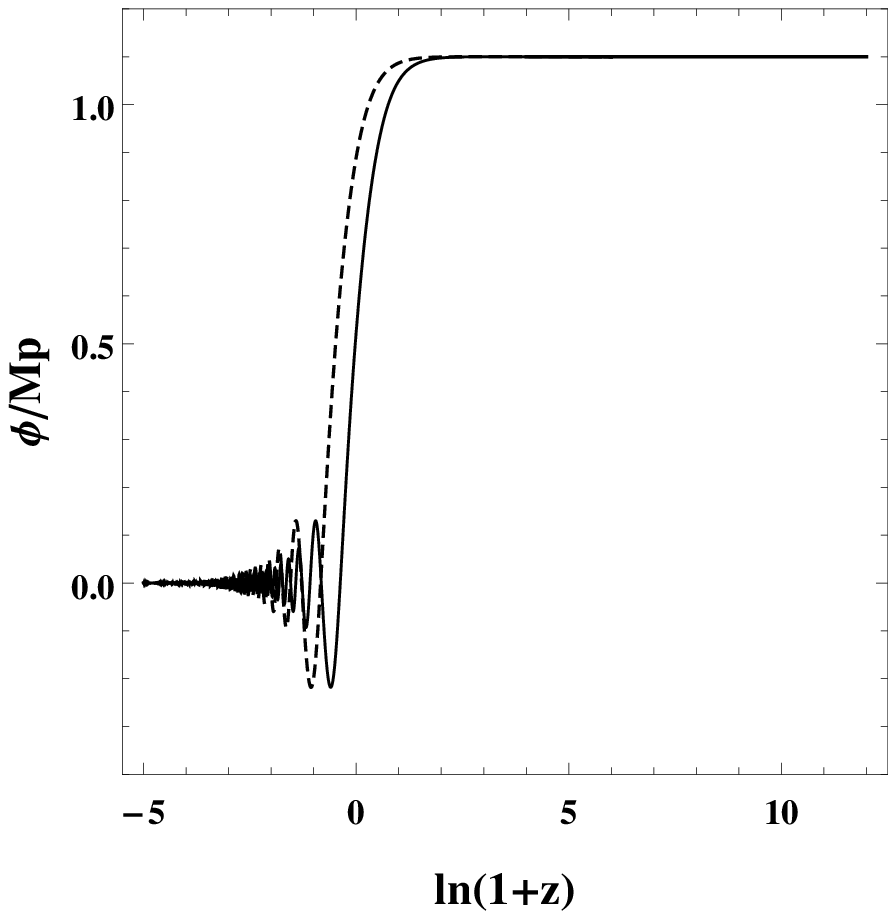}} &
{\includegraphics[width=1.6in,height=1.6in,angle=0]{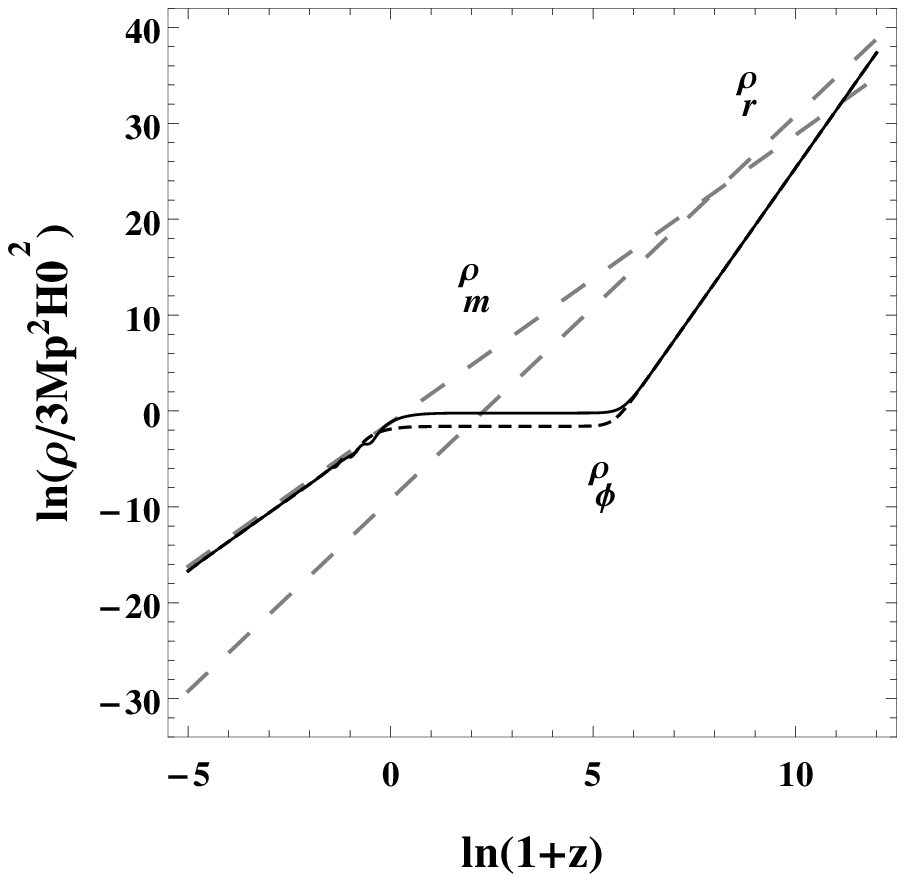}} 
 \\
{\includegraphics[width=1.6in,height=1.6in,angle=0]{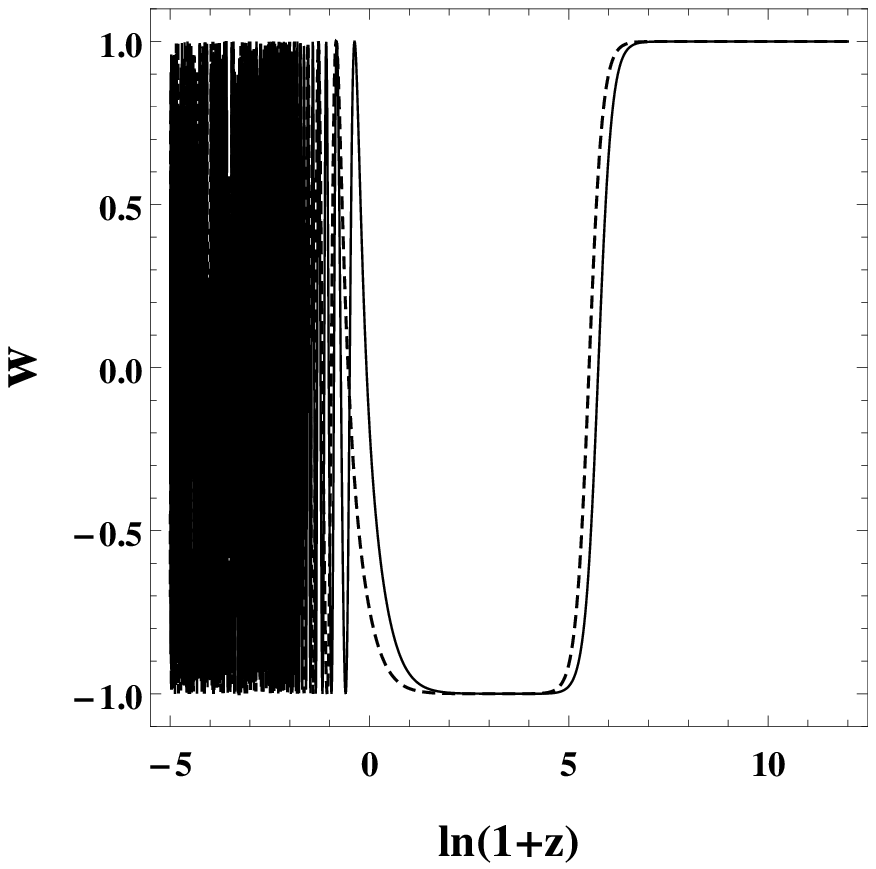}} &
{\includegraphics[width=1.6in,height=1.6in,angle=0]{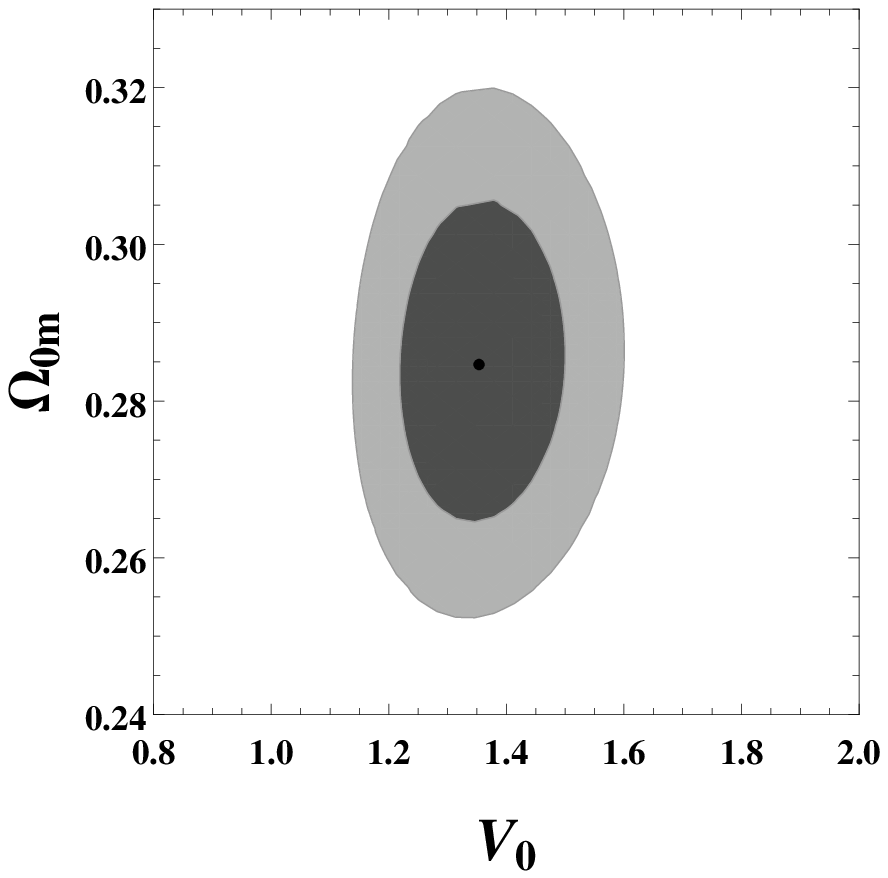}}
\end{tabular}
\end{center}
\caption{This figure is plotted for the quadratic potential (\ref{eq:quad}), and exhibits the evolution of the scalar field $\phi$, energy density $\rho$ and equation of state $w$ versus the redshift $z$. The dashed and solid lines correspond to the evolution of $\phi$, $\rho_{\phi}$ and $w$ for two different values of $V_0=0.5 M_p^2 H_0^2$ (dashed) and $V_0=2  M_p^2 H_0^2$ (solid). The big dashed ($-$) lines represent the evolution of the energy densities of matter and radiation. The bottom right panel shows 1$\sigma$ (dark shaded) and 2$\sigma$ (light shaded) likelihood contours. We have used the joint data (SN+Hubble+BAO+CMB) to carry out the data analysis. The black dot designates the best-fit value of the model parameters, and found to be $V_0=1.3534$ and $\Omega_{0m}=0.2846$.}
\label{figrhoq}
\end{figure}

Our study shows that the $\alpha$ attractor model behaves as a $\phi^2$ potential when the scalar field approaches to origin. Before reaching the oscillatory phase, it has a slow-roll regime, which provides the late-time cosmic acceleration at the present epoch. Similar results can be obtained,  if we use the $\phi^2$ potential rather than the $\alpha$ attractor. To confirm this, we consider the quadratic potential
\begin{eqnarray}
V(\phi) = V_0 \left(\frac{\phi}{M_p}\right)^2
\label{eq:quad}
\end{eqnarray}
We numerically evolve Eqs.(\ref{evol1d}) and (\ref{evol2d}) with (\ref{eq:quad}). The results are presented in figure \ref{figrhoq}. The top and bottom left panels show the evolution of $\phi$, $\rho_{\phi}$ and $w$ versus the redshift. One can clearly see that before reaching the oscillatory phase, the field has a slow-roll regime, which gives the late-time cosmic acceleration at the current epoch. The bottom right panel exhibits the 1$\sigma$ (dark shaded) and 2$\sigma$ (light shaded) likelihood contours in the $V_0 - \Omega_{0m}$ plane. The best-fit values of the model parameters are found to be $V_0=1.3534$ and $\Omega_{0m}=0.2846$. The best-fit values of $\Omega_{0m}$ for the quadratic potential and $\alpha$ model almost coincide, see Table \ref{bestt}. Therefore,  both  models provide a similar  late-time cosmic acceleration at the present epoch, and are consistent with current observations.
%%%%%%%%%
%%%%%%%
\section{Data Analysis}
\label{DA}
We employ the $\chi^2$ analysis to constrain the model parameters. One can use the maximum likelihood method and get the total likelihood for the parameters $\alpha$, $\Omega_{0m}$ and $c$ as the product of individual likelihood for different datasets. The total likelihood function for joint data is given by  
\begin{equation}
\mathcal{L}_{tot}(\alpha, \Omega_{0m}, c) = e^{- \frac{\chi_{tot}^2(\alpha, \Omega_{0m}, c)}{2}}
\end{equation}
where
\begin{align}
\chi_{\rm tot}^2=\chi_{\rm SN}^2+\chi_{\rm Hub}^2+\chi_{\rm BAO}^2+\chi_{\rm CMB}^2 \,,
\label{chi_tot}
\end{align}
is related to Type Ia supernova (SN), data of Hubble parameter, Baryon Acoustic Oscillation (BAO) and Cosmic Microwave Background (CMB) data. The best fit value of the parameters is acquired by minimizing $\chi_{\rm tot}^2$ with respect to $\alpha$, $\Omega_{0m}$  and $c$. The likelihood contours in 1$\sigma$ and 2$\sigma$ confidence level are given as 2.3 and 6.17, respectively, in the two dimensional parametric space.
\subsection{Type Ia Supernova}
\label{SN}
To study the Universe on a very large scale, Type Ia supernova is considered as an ideal astronomical object. They are very bright and the luminosity distance can be determined upto the  redshift $z \simeq 1.4$. They have almost the same luminosity which is  redshift-independent. Hence Type Ia supernova are observed as very good standard candles. 

The Type Ia supernova is one of the direct probe for the cosmological expansion. We take 580 data points of latest Union2.1 compilation data \cite{Suzuki:2011hu}. In this case, one measures the apparent luminosity of the supernova explosion. The most appropriate quantity is the luminosity distance $D_L(z)$ defined as
\begin{equation}
D_L(z)=(1+z)
\int_0^z\frac{H_0dz'}{H(z')}.
\end{equation}
In reality, the distance modulus $\mu(z)$ is the observed quantity which is directly related to the $D_L(z)$ as  $\mu(z)=m - M=5 \log D_L(z)+\mu_0$, where $M$ and $m$
are the absolute and apparent magnitudes of the Supernovae and
$\mu_0=5 \log\left(\frac{H_0^{-1}}{\rm Mpc}\right)+2 5$ is a
nuisance parameter which should be marginalized. Therefore, the corresponding $\chi^2$ can be written as
\begin{align}
\chi_{\rm SN}^2(\mu_0,\theta)=\sum_{i=1}^{580}
\frac{\left[\mu_{th}(z_i,\mu_0,\theta)-\mu_{obs}(z_i)\right]^2}{
\sigma_\mu(z_i)^2}\,,
\end{align}
where $\mu_{obs}$, $\mu_{th}$  and $\sigma_{\mu}$ represent the observed, theoretical distance modulus and uncertainty in the distance modulus respectively; $\theta$ represents any arbitrary parameter of the particular model. Eventually marginalizing $\mu_0$  and following reference \cite{Lazkoz05}, we get
\begin{align}
\chi_{\rm SN}^2(\theta)=A(\theta)-\frac{B(\theta)^2}{C(\theta)}\,,
\end{align}
where,
\begin{align}
&A(\theta) =\sum_{i=1}^{580}
\frac{\left[\mu_{th}(z_i,\mu_0=0,\theta)-\mu_{obs}(z_i)\right]^2}{
\sigma_\mu(z_i)^2}\,, \\
%\end{align}
%\begin{align}
&B(\theta) =\sum_{i=1}^{580}
\frac{\mu_{th}(z_i,\mu_0=0,\theta)-\mu_{obs}(z_i)}{\sigma_\mu(z_i)^2}\,, \\
&C(\theta) =\sum_{i=1}^{580} \frac{1}{\sigma_\mu(z_i)^2}\,.
\end{align}
\subsection{The Hubble Parameter $H(z)$}
\label{Hub}
The Hubble Parameter $H(z)$ represents the expansion history of the Universe and plays a key role in joining the cosmological models and observations. Recently, Farooq and Ratra \cite{Farooq:2013hq} compiled 28 data points for $H(z)$ in the redshift range $0.07 \leq z \leq 2.3$ which are given in Table \ref{hubt}. To complete the data set we take $H_0 = 67.8 \pm 0.9~ Km/S/M_{Pc}$ from Planck results \cite{plank data}. We shall work with the normalized Hubble parameter, $h = H/H_0$ and apply the data to the model. In this case, the $\chi^2$ is defined as 
\begin{align}
\chi_{\rm Hub}^2(\theta)=\sum_{i=1}^{29}
\frac{\left[h_{\rm th}(z_i,\theta)-h_{\rm obs}(z_i)\right]^2}{
\sigma_h(z_i)^2}\,,
\end{align}
where $h_{\rm obs}$ and $h_{\rm th}$ are respectively the  observed and theoretical values of the normalized Hubble parameter. Also,
\be
\sigma_h = \left( \frac{\sigma_H}{H}+\frac{\sigma_{H_0}}{H_0} \right) h,
\label{errorh}
\ee
where $\sigma_H$  and $\sigma_{H_0}$ are the errors associated with $H$ and ${H_0}$ respectively.
%%%%%%%%%
\begin{table}
\caption{$H(z)$ measurements (in unit [$\mathrm{km\,s^{-1}Mpc^{-1}}$]) and their errors \cite{Farooq:2013hq}.}
\begin{center}
\label{hubt}
\begin{tabular}{cccc}
\hline\hline
~~$z$ & ~~~~$H(z)$ &~~~~ $\sigma_{H}$ & ~~ Reference\\
~~    &~~ (km/s/Mpc) &~~~~ (km/s/Mpc)& \\
\tableline
0.070&~~    69&~~~~~~~  19.6&~~ \cite{Zhang:2012mp}\\
0.100&~~    69&~~~~~~~  12&~~   \cite{Simon:2004tf}\\
0.120&~~    68.6&~~~~~~~    26.2&~~ \cite{Zhang:2012mp}\\
0.170&~~    83&~~~~~~~  8&~~    \cite{Simon:2004tf}\\
0.179&~~    75&~~~~~~~  4&~~    \cite{Moresco:2012by}\\
0.199&~~    75&~~~~~~~  5&~~    \cite{Moresco:2012by}\\
0.200&~~    72.9&~~~~~~~    29.6&~~ \cite{Zhang:2012mp}\\
0.270&~~    77&~~~~~~~  14&~~   \cite{Simon:2004tf}\\
0.280&~~    88.8&~~~~~~~    36.6&~~ \cite{Zhang:2012mp}\\
0.350&~~    76.3&~~~~~~~    5.6&~~  \cite{Chuang2012b}\\
0.352&~~    83&~~~~~~~  14&~~   \cite{Moresco:2012by}\\
0.400&~~    95&~~~~~~~  17&~~   \cite{Simon:2004tf}\\
0.440&~~    82.6&~~~~~~~    7.8&~~  \cite{Blake12}\\
0.480&~~    97&~~~~~~~  62&~~   \cite{Stern:2009ep}\\
0.593&~~    104&~~~~~~~ 13&~~   \cite{Moresco:2012by}\\
0.600&~~    87.9&~~~~~~~    6.1&~~  \cite{Blake12}\\
0.680&~~    92&~~~~~~~  8&~~    \cite{Moresco:2012by}\\
0.730&~~    97.3&~~~~~~~    7.0&~~  \cite{Blake12}\\
0.781&~~    105&~~~~~~~ 12&~~   \cite{Moresco:2012by}\\
0.875&~~    125&~~~~~~~ 17&~~   \cite{Moresco:2012by}\\
0.880&~~    90&~~~~~~~  40&~~   \cite{Stern:2009ep}\\
0.900&~~    117&~~~~~~~ 23&~~   \cite{Simon:2004tf}\\
1.037&~~    154&~~~~~~~ 20&~~   \cite{Moresco:2012by}\\
1.300&~~    168&~~~~~~~ 17&~~   \cite{Simon:2004tf}\\
1.430&~~    177&~~~~~~~ 18&~~   \cite{Simon:2004tf}\\
1.530&~~    140&~~~~~~~ 14&~~   \cite{Simon:2004tf}\\
1.750&~~    202&~~~~~~~ 40&~~   \cite{Simon:2004tf}\\
2.300&~~    224&~~~~~~~ 8&~~    \cite{Busca12}\\

\hline\hline
\end{tabular}
\end{center}
\end{table}
%%%%%%%
%%%%%%%%%%%%
\subsection{Baryon Acoustic Oscillation (BAO)}
\label{BAO} 
The early Universe composed of photons, baryons and dark matter. Photons and baryons are tightly coupled to one another through Thompson scattering, and act as a single fluid. This fluid can not collapse under gravity but it can oscillate, due to the large pressure furnished by the photons. These oscillations are known as BAO, which are the consequences of photon-baryon coupling at redshift larger than $z=1090$.

The characteristic scale of these oscillations is governed by the sound horizon $r_s$ at the photon decoupling epoch, given as:
\begin{eqnarray}
r_{s} (z_{*}) &=& \frac{c}{\sqrt{3}} \int_{0}^{\frac{1}{1+z_{*}}} \frac{da}{a^2 H(a) \sqrt{1+(3 \Omega_{0b}/4\Omega_{0\gamma})a}},
\label{eq:rs}
\end{eqnarray} 
where $\Omega_{0b}$ and $\Omega_{0\gamma}$ are the present values of baryon and photon density parameter respectively, and $z_{*}$ is the redshift of photon decoupling.

The BAO sound horizon scale can be used to derive the angular diameter distance $D_A$ and the Hubble expansion rate $H$ as a function of redshift. By measuring the subtended angle $\Delta \theta$, of the ruler of length $r_{s}$, these parameters are defined as follows:
\begin{eqnarray}
\Delta \theta &=& \frac{r_{s}}{d_{A}{(z)}} \qquad \mbox{with} \qquad d_A(z)= \int_0^z\frac{dz'}{H(z')}
\label{eq:theta}
\end{eqnarray} 
where $\Delta \theta$ is the measured angular separation of the BAO feature in the 2 point correlation function of the galaxy distribution on the sky, and  
\begin{eqnarray}
\Delta z &=& H(z) r_s ,
\label{eq:Hrs}
\end{eqnarray} 
where $\Delta z$ is the measured redshift separation of the BAO feature in the 2 point correlation function along the line of sight. We work with BAO data of $d_A(z_\star)/D_V(Z_{BAO})$ \cite{Blake:2011en, Percival:2009xn, Beutler:2011hx, Jarosik:2010iu, Eisenstein:2005su, Giostri:2012ek}, where $z_\star \approx 1091$ is the decoupling time,  $d_A(z)$ is the co-moving angular-diameter distance and  $D_V(z)=\left(d_A(z)^2 z/H(z)\right)^{1/3}$ is the dilation scale. Data needed for this inspection is shown in Table \ref{BAOt}. The corresponding $\chi_\mathrm{BAO}^2$ is given as \cite{Giostri:2012ek}:
\begin{equation}
\chi_{\rm BAO}^2=X^T C^{-1} X\,,
\end{equation}
where
\begin{equation}
X=\left( \begin{array}{c}
        \frac{d_A(z_\star)}{D_V(0.106)} - 30.95 \\
        \frac{d_A(z_\star)}{D_V(0.2)} - 17.55 \\
        \frac{d_A(z_\star)}{D_V(0.35)} - 10.11 \\
        \frac{d_A(z_\star)}{D_V(0.44)} - 8.44 \\
        \frac{d_A(z_\star)}{D_V(0.6)} - 6.69 \\
        \frac{d_A(z_\star)}{D_V(0.73)} - 5.45
        \end{array} \right)\,,
\end{equation}
and $C^{-1}$ is the inverse covariance matrix defined as in \cite{Giostri:2012ek}.

\begin{table*}
\caption{Values of $\frac{d_A(z_\star)}{D_V(Z_{BAO})}$ for distinct values of $z_{BAO}$.}
\begin{center}
%\resizebox{\textwidth}{!}{%
\begin{tabular}{c||cccccc}
\hline\hline
~~~~~~~~~$z_{BAO}$~~  & ~~0.106~  & 0.2~& 0.35~ & 0.44~& 0.6~& 0.73~~\\
\hline
~~~~~~~~~ $\frac{d_A(z_\star)}{D_V(Z_{BAO})}$~~ &  ~~$30.95 \pm 1.46$~~~ & $17.55 \pm 0.60$~~~
& $10.11 \pm 0.37$ ~~~& $8.44 \pm 0.67$~~~ & $6.69 \pm 0.33$~~~ & $5.45 \pm 0.31$~~
\\
\hline\hline
\end{tabular}
%}
\label{BAOt}
\end{center}
\end{table*}
\begin{align}
C^{-1}=\left(
\begin{array}{cccccc}
 0.48435 & -0.101383 & -0.164945 & -0.0305703 & -0.097874 & -0.106738 \\
 -0.101383 & 3.2882 & -2.45497 & -0.0787898 & -0.252254 & -0.2751 \\
 -0.164945 & -2.45499 & 9.55916 & -0.128187 & -0.410404 & -0.447574 \\
 -0.0305703 & -0.0787898 & -0.128187 & 2.78728 & -2.75632 & 1.16437 \\
 -0.097874 & -0.252254 & -0.410404 & -2.75632 & 14.9245 & -7.32441 \\
 -0.106738 & -0.2751 & -0.447574 & 1.16437 & -7.32441 & 14.5022
\end{array}
\right)\,.
\end{align}
\subsection{Cosmic Microwave Background (CMB) distance information}
\label{CMB} 
The CMB measurement is sensitive to distance to the last scattering surface (decoupling epoch) via the positions of peaks and troughs of acoustic oscillations. Following the WMAP results  \cite{komatsu}, the distance information incorporates the ``shift parameter'' $R$, ``acoustic scale'' $l_A$ and the redshift of last scattering surface $z_{ls}$, where $R$ and $l_A$ are the ratio of angular diameter distance to the last scattering surface epoch over the Hubble horizon and the sound horizon at surface of the last scattering, and are given by
\begin{eqnarray}
R &=&  H_{0} \sqrt{ \Omega_{0m}} ~\chi (z_{ls}),\\
l_{A} &=& \frac{\pi \chi (z_{ls})}{\chi_{s}(z_{ls})},
\end{eqnarray} 
where $\chi (z_{ls})$ is the co-moving distance to $z_{ls}$ and $\chi_{s} (z_{ls})$ is the co-moving sound horizon at $z_{ls}$. The shift parameter $R$ can also be computed theoretically using the formula
\begin{eqnarray}
R &=&  H_{0} \sqrt{ \Omega_{0m}} \int_{0}^{z_{ls}} \frac{dz'}{H(z')}.
\label{eq:R}
\end{eqnarray} 
From equation (\ref{eq:R}), we observe that $R$ is related to the matter density as well as the expansion history of the Universe until the redshift of the surface of  the last scattering, $z_{ls}$,  which is computed through the fitting function \cite{Hu}:
\begin{eqnarray}
z_{ls} &=&  1048 \left[1+0.00124 (\Omega_{b} h^2)^{-0.738} \right] \left[1+g_1 (\Omega_{0m} h^2)^{g_2} \right],
\label{eq:zls}
\end{eqnarray} 
where $g_1$ and $g_2$ are defined as
\begin{eqnarray}
g_{1} &=&  \frac{0.0783 (\Omega_{b} h^2)^{-0.238}}{1+39.5 (\Omega_{b} h^2)^{0.763}},\\
g_{2} &=&  \frac{0.56}{1+21.1 (\Omega_{b} h^2)^{1.81}}.
\label{eq:g1g2}
\end{eqnarray} 
The corresponding $\chi^2$ can be written as
\begin{align}
\chi^2(\theta)=\frac{(R(\theta)-R_0)^2}{\sigma^2}\,,
\end{align}
where $R(\theta)$ depends  on the model parameter $\theta$ and $R_0=1.725 \pm 0.018$ \cite{komatsu}.

\section{Conclusion}
\label{conclusion}
In this paper, we have investigated the generalized $\alpha$ attractor model that leads to the cosmological attractor behavior and can interpolate between the thawing and scaling freezing models. At  the present epoch, it behaves as a thawing model whereas in the future it possesses scaling freezing behavior. In Figure \ref{figrho}, the dynamics of the scalar field and  energy density $\rho_{\phi}$ are shown for $\alpha=0.05, 0.1, 1, 10$. The evolution of $\rho_{\phi}$ starts off as a thawing dark energy along with thawing behavior, and even at  the present epoch it lies in the thawing region. But in the future it switches over and turns to the scaling regime, which is cosmological attractor. For small values of  the scalar field $\phi$, the  generalized $\alpha$ attractor model mimics the power law behavior $V(\phi)\sim \phi^2$ and produces oscillations of $\phi$ near the origin which are reflected in the behavior of the equation of state $w$ (see Figure \ref{figrho}), and correspondingly the average equation of state parameter is $ < w > =0$ (see Figure \ref{figwavg}). In our setting, the scalar field remains in the slow roll regime but mimics the scaling behavior in the future. We considered two cases $n= 1, 2$, and used the joint data (SN+Hubble+BAO+CMB) to obtain the observational constraints on the model parameters. The best-fitting values of the model parameters for $n=1, 2$ are  $\alpha=0.0715$, $\Omega_{0m}=0.2841$ and $c=11.4884$, and $\alpha=0.0720$, $\Omega_{0m}=0.2849$ and $c=22.0320$, respectively.

We also considered the quadratic potential in order to compare with the results obtained from the $\alpha$ model. Figure \ref{figrhoq} showed the evolution of $\phi$, $\rho_{\phi}$ and $w$ versus the redshift, and the 1$\sigma$ (dark shaded) and 2$\sigma$ (light shaded) likelihood contours in the $V_0 - \Omega_{0m}$ plane. The best-fit values of the model parameters are found to be $V_0=1.3534$ and $\Omega_{0m}=0.2846$. The best-fit values of $\Omega_{0m}$ for both  models are almost the same. Hence, in view of the current observations, both models are viable, and provide  the late-time cosmic acceleration at the present epoch.
%%%%%%%%%
\section*{Acknowledgements}
M.S. thanks M. Sami for making useful comments and suggestions. He is also thankful to M. Sajjad Athar for his constant encouragement throughout the work. A.W. is supported in part by NNSFC No.11375153 and No. 11675145, China.
%%%%%

\end{document}